\def\presentation{
\voffset -.50in  \hoffset -.19in
\oddsidemargin 0in \evensidemargin 0in
\marginparwidth .75in \marginparsep 7pt \topmargin 0in
\headheight 12pt \headsep .25in
\footheight 18pt \footskip .35in
\textheight 9.5in \textwidth 6.5in
\columnsep 10pt \columnseprule 0pt }
\documentstyle[10pt]{article}
\input{epsf}
\presentation
\begin{document}
%

%
\def\tilde{\widetilde}
\def\bar{\overline}
\def\hat{\widehat}
\def\*{\star}
\def\[{\left[}
\def\]{\right]}
\def\({\left(}
\def\){\right)}
\def\zb{{\bar{z} }}
\def\frac#1#2{{#1 \over #2}}
\def\inv#1{{1 \over #1}}
\def\half{{1 \over 2}}
\def\d{\partial}
\def\der#1{{\partial \over \partial #1}}
\def\dd#1#2{{\partial #1 \over \partial #2}}
\def\vev#1{\langle #1 \rangle}
\def\bra#1{{\langle #1 |  }}
\def\ket#1{ | #1 \rangle}
\def\rvac{\hbox{$\vert 0\rangle$}}
\def\lvac{\hbox{$\langle 0 \vert $}}
\def\2pi{\hbox{$2\pi i$}}
\def\e#1{{\rm e}^{^{\textstyle #1}}}
\def\grad#1{\,\nabla\!_{{#1}}\,}
\def\dsl{\raise.15ex\hbox{/}\kern-.57em\partial}
\def\Dsl{\,\raise.15ex\hbox{/}\mkern-.13.5mu D}
\def\comm#1#2{ \BBL\ #1\ ,\ #2 \BBR }
\def\x{\stackrel{\otimes}{,}}
\def\det{ {\rm det}}
\def\tr{{\rm tr}}
%
%
\def\th{\theta}         \def\Th{\Theta}
\def\ga{\gamma}         \def\Ga{\Gamma}
\def\be{\beta}
\def\al{\alpha}
\def\ep{\epsilon}
\def\la{\lambda}        \def\La{\Lambda}
\def\de{\delta}         \def\De{\Delta}
\def\om{\omega}         \def\Om{\Omega}
\def\sig{\sigma}        \def\Sig{\Sigma}
\def\vphi{\varphi}      \def\ee{{\rm e}}
%
%
\def\CA{{\cal A}}       \def\CB{{\cal B}}       \def\CC{{\cal C}}
\def\CD{{\cal D}}       \def\CE{{\cal E}}       \def\CF{{\cal F}}
\def\CG{{\cal G}}       \def\CH{{\cal H}}       \def\CI{{\cal J}}
\def\CJ{{\cal J}}       \def\CK{{\cal K}}       \def\CL{{\cal L}}
\def\CM{{\cal M}}       \def\CN{{\cal N}}       \def\CO{{\cal O}}
\def\CP{{\cal P}}       \def\CQ{{\cal Q}}       \def\CR{{\cal R}}
\def\CS{{\cal S}}       \def\CT{{\cal T}}       \def\CU{{\cal U}}
\def\CV{{\cal V}}       \def\CW{{\cal W}}       \def\CX{{\cal X}}
\def\CY{{\cal Y}}       \def\CZ{{\cal Z}}
%
%
\font\numbers=cmss12
\font\upright=cmu10 scaled\magstep1
\def\stroke{\vrule height8pt width0.4pt depth-0.1pt}
\def\topfleck{\vrule height8pt width0.5pt depth-5.9pt}
\def\botfleck{\vrule height2pt width0.5pt depth0.1pt}
\def\Zmath{\vcenter{\hbox{\numbers\rlap{\rlap{Z}\kern
		0.8pt\topfleck}\kern
		2.2pt \rlap Z\kern 6pt\botfleck\kern 1pt}}}
\def\Qmath{\vcenter{\hbox{\upright\rlap{\rlap{Q}\kern
		   3.8pt\stroke}\phantom{Q}}}}
\def\Nmath{\vcenter{\hbox{\upright\rlap{I}\kern 1.7pt N}}}
\def\Cmath{\vcenter{\hbox{\upright\rlap{\rlap{C}\kern
		   3.8pt\stroke}\phantom{C}}}}
\def\Rmath{\vcenter{\hbox{\upright\rlap{I}\kern 1.7pt R}}}
\def\Z{\ifmmode\Zmath\else$\Zmath$\fi}
\def\Q{\ifmmode\Qmath\else$\Qmath$\fi}
\def\N{\ifmmode\Nmath\else$\Nmath$\fi}
\def\C{\ifmmode\Cmath\else$\Cmath$\fi}
\def\R{\ifmmode\Rmath\else$\Rmath$\fi}
\def\cadremath#1{\vbox{\hrule\hbox{\vrule\kern8pt\vbox{\kern8pt
			\hbox{$\displaystyle #1$}\kern8pt} 
			\kern8pt\vrule}\hrule}}
\def\proof{\noindent {\underline {Proof}.} }
\def\cqfd{ {\hfill{$\Box$}} }
\def\square{ {\hfill \vrule height6pt width6pt depth1pt} } 
%
%
\def\debut{ \begin{eqnarray} }
\def\fin{ \end{eqnarray} }
\def\non{ \nonumber }
\def\bu{ \bar{u} }
%

%
%
\rightline{SPhT/98/136}
\vskip 2cm
\centerline{\LARGE\bf Sailing the Deep Blue Sea}
\vskip 0.2cm
\centerline{\LARGE\bf of Decaying Burgers Turbulence 
\footnote[0]{or, {\bf ``Lagrangian Trajectories in
Decaying Burgers Turbulence"} .}.}
\vskip 1.5cm

\centerline{\large  
Michel Bauer\footnote[1]{bauer@spht.saclay.cea.fr} and 
Denis Bernard\footnote[2]{Membre 
du CNRS; dbernard@spht.saclay.cea.fr} 
}
\bigskip 
\centerline{Service de Physique Th\'eorique 
de Saclay\footnote[3]{Laboratoire de la Direction des Sciences 
de la Mati\`ere du Commissariat \`a l'Energie Atomique.}}
\centerline{F-91191, Gif-sur-Yvette, France.}
\medskip

\vskip2.5cm

\centerline{\bf{Abstract}}
We study Lagrangian trajectories and scalar transport statistics
in decaying Burgers turbulence. We choose velocity fields 
solutions of the inviscid Burgers equation whose probability
distributions are specified by Kida's statistics. They are 
time-correlated, not time-reversal invariant and not Gaussian.
We discuss in some details the effect of shocks on trajectories
and transport equations. We derive the inviscid
limit of these equations using a formalism of operators
localized on shocks. We compute the probability distribution
functions of the trajectories although they do not define Markov
processes. As physically expected, these trajectories are
statistically well-defined but collapse with probability
one at infinite time. We point out that the
advected scalars enjoy inverse energy cascades. We also make a few
comments on the connection between our computations and persistence
problems. 

\vfill
\newpage
 

\section{Introduction}
Lagrangian trajectories driven by a velocity field $u(x,t)$
are solutions of the differential equation:
\debut
\frac{dx(t)}{dt} = u(x(t),t) \label{lagran}
\fin
As known from the precursors, Richardson, Kolmogorov, Batchelor,... \cite{ancien},
they acquire peculiar properties when the flow becomes turbulent.
These properties are probably going to play an important role in
the understanding of fully developed turbulence. For example, 
the recent proof \cite{sinaietal} of the existence and
uniqueness  of the stationary state for the inviscid forced Burgers
turbulence is based on an analysis of these trajectories.

Statistical properties of these trajectories may be 
deciphered by looking at transport phenomena in turbulent systems.
Recent studies  of the Kraichnan's advection models \cite{krach} 
have made these expected properties more
explicit. Kraichnan models assume that the velocity
fields is Gaussian and white-noise in time. These
simplifications lead to the solvability 
of the models. See refs.\cite{incompress} for recent studies of
the Kraichnan models for incompressible fluids
and refs.\cite{compress,GaVer} for compressible ones.
Two kinds of behavior have been observed:\\
(1) Statistical ill-definedness, meaning that two trajectories
starting at the same point have a non vanishing probability
to be far apart at later time,\\
(2) Trajectory collapse for compressible enough fluids,
meaning that two trajectories starting initially 
at different positions have a non-zero probability
to follow the same path after some time,\\
However:\\
(3) Properties (1) and (2) do not seem to occur simultaneoulsy.
\vskip 0.2 cm

The motivation of the present work is to decipher whether 
these properties are more robust and hold true  
for more realistic velocity fields than those chosen
in Kraichnan's models. 
Of course we could not solve the problem with a 
velocity field describing a real three dimensional turbulent system. 
Instead we shall consider (unrealistic) velocity fields, 
solutions of the Burgers equation which in 1+1 dimensions
takes the form:
\debut
\d_t u + u \d_x u - \nu \d_x^2 u = 0 \label{burgers}
\fin
where $u= u(x,t)$ is the (compressible) velocity field and $\nu$ the viscosity. 
This is a variant of the Navier-Stokes equation in which the role 
of the pressure has been neglected.
Although we shall stick to one dimensional space, 
some of the following considerations could be generalized 
to higher dimensions.

No external force is applied to equation (\ref{burgers}).
So its inviscid limit $\nu \to 0$
corresponds to decaying turbulence whose statistical  
description consists in finding probability
distribution of the velocity fields solution of eq.(\ref{burgers})
given random initial data. One usually expects a more
universal behavior at large time.  
Thus, we shall consider a family of velocity fields,
solutions of the inviscid limit $\nu\to 0$ of the Burgers equation,
whose probability distribution  describes the long time
behavior of large classes of initial conditions. 
These velocity statistics are those first introduced by 
Kida \cite{kida}. In contrast to the Kraichnan model,
the velocity fields are then  not white-noise in time, 
not time-reversal invariant and not Gaussian.
\vskip 0.2 cm

For compressible fluids, one may look at two kinds of transport
phenomena depending on whether one is looking at the advection
of a tracer, that we shall denote by $T(x,t)$, or
at the advection of the density of a pollutant, that we
shall denote by $\rho(x,t)$. The corresponding viscosity is written
$\kappa$. The equations
governing these transports are: 
\debut
\d_t T(x,t) + u(x,t) \d_x T(x,t) -\kappa \d_x^2 T(x,t) = 0 \label{pourT}\\
\d_t \rho(x,t) + \d_x(u(x,t) \rho(x,t))
-\kappa \d_x^2 \rho(x,t) = 0\label{pourrho}
\fin
They differ by the order of the derivative and velocity.

In the inviscid limit $\nu\to 0$, solutions of the Burgers
equation develop shocks at which the velocity is not smooth.
This non-smoothness implies that the naive definition of
the trajectories does not apply. Therefore, these trajectories
and the transport equations have to be dealt with carefully.
As we shall see, a correct definition of the transport
equations will turn out to be:
\debut
\d_t T(x,t) + \half\Bigl({u(x^+,t)+u(x^-,t)}\Bigr) \d_x T(x,t) 
-\kappa \d_x^2 T(x,t) = 0 \label{Tbis}\\
\d_t \rho(x,t) + \d_x\half\Bigl(\Bigl({u(x^+,t)+u(x^-,t)}\Bigr)\rho(x,t)\Bigr) 
-\kappa \d_x^2 \rho(x,t) = 0 \label{rhobis}
\fin
with $u(x^\pm,t)=\lim_{\ep\to 0^+}u(x\pm\ep,t)$.
Although equations (\ref{Tbis},\ref{rhobis}) seem to be naively equivalent
to equations (\ref{pourT},\ref{pourrho}), they are not since in
the inviscid limit the velocity field $u(x,t)$ is not smooth.

In the limit $\kappa\to 0$, eqs.(\ref{Tbis},\ref{rhobis})
have a natural interpretation in terms of Lagrangian
trajectories. However, the naive equation (\ref{lagran}), 
which is actually meaningless since $u(x,t)$ is discontinuous,
has to be modified into:
\debut
\frac{dx(t)}{dt}\vert_+ = \half\Bigl({ u(x(t)^+,t) + u(x(t)^-,t) }\Bigr) \label{lagun}
\fin
Again this differs from eq.(\ref{lagran}) because
$u(x,t)$ is not smooth. The physical meaning of this modification is
clear. At points of 
discontinuity, the ill-defined velocity is replaced by the velocity of
the shock which, as is well known since long ago \cite{burg, kida,
compars}, is just the average
of the velocities just before and just after the shock. 
Once this will be done, we shall describe how to compute
the probability distribution functions (p.d.f.) of the trajectories 
and we shall use them to discuss the properties
of the transport equation (\ref{Tbis}) in the limit $\kappa\to 0$.
\vskip 0.2cm

This paper is organized as follows.
In the following section we recall basic facts concerning
the Burgers equation and the velocity profiles we shall use.
Section 3 is devoted to give a precise the definition of
Lagrangian trajectories in the inviscid limit $\nu \to 0$ and to the
relation with the correct form of transport equations and their solutions.
In Section 4 we establish identities, called equations of motion,
which are valid inside correlation functions. This is based on
operators localized at shocks and their algebra.
In Section 5, the backward and forward probability distribution functions of
the trajectories are introduced and their formal properties
emphasized. We use the identities established in Section 4 to 
verify that these p.d.f. are solutions of the transport equations.
In Section 6, we make explicit computations for one and two particle
distributions. We check the consistency with the expected physical
properties of the trajectories. In particular, we show by different
approaches that the trajectories 
are statistically well-defined but that particles have a non-vanishing
probability to collapse. This is in agreement with the general
properties of Lagrangian trajectories mentioned above as (1),(2) and
(3): As we deal with a highly compressible fluid, the alternative (2) is
realized and the alternative (1) is excluded. The connexion with persistence
problems is made. Finally arguments indicating that the
energy cascade in scalar advection in these flows is inverse,
i.e. towards the large scale, are presented in Section 7.
\vskip 0.2cm

\section{Velocity profiles}

This short section is devoted to the specification of the statistics
of the velocity profiles to be used in this paper.

$\bullet$ In order to fix notations, we recall a few elementary 
facts concerning Burgers equation (see e.g. \cite{burg,kida} 
and \cite{compars} and references therein).
As is well known, the equation is solved by implementing the Cole-Hopf 
transformation which maps it to the heat equation. 
This works as follows.
Let $Z(x,t)=\exp[-\inv{2\nu}\Phi(x,t)]$ where $u(x,t)=\d_x \Phi(x,t)$.
Eq.\,(\ref{burgers}) for $u$ is mapped into the heat equation for $Z$:
\debut
\[{\d_t - \nu \d_x^2}\] Z(x,t)=0\,. \non
\fin
Thus, given the initial condition $u(x,t=0)\equiv u_0(x)$, the velocity
field at a later time $t$ is recovered from the potential $\Phi(x,t)$
given by the relation
\debut
\exp\[{-\inv{2\nu}\Phi(x,t)}\] = \int \frac{dy}{\sqrt{4\pi \nu t}}
\exp\[{ -\inv{2\nu}\({\Phi_0(y) +\frac{(x-y)^2}{2t} }\) }\] \label{solu}
\fin
with $\Phi_0(x)$ standing for the initial potential such that 
$u_0(x)=\d_x\Phi_0(x)$. The inviscid Burgers equation  corresponds 
to the limit $\nu \to 0$. The solution is then given by solving 
a minimalization problem: 
\debut
u(x,t) = \d_x\Phi(x,t) \quad {\rm with}\quad
\Phi(x,t)= \min_y \({\Phi_0(y) +\frac{(x-y)^2}{2t} }\). \label{minsol}
\fin
Outside shocks the minimum is reached for only one value $y_*$ of $y$,
the solution of the equation $u_0(y_*)t=x-y_*$. The velocity is 
$u(x,t)=\frac{x-y_*}{t}=u_0(y_*)$. It is effectively a local solution
of the inviscid Burgers equation since, by the minimum condition
defining $y_*$, we have $u(x,t)= u_0(x- t u(x,t))$. 
A simple geometrical construction of the solution (\ref{minsol}) 
is described in refs.\cite{burg,kida}. For large $t$, $y_*$ coincides 
approximately with one of the local minima of $\Phi_0(y)$ and
it practically does not change under small variations of $x$
so that, in between the shocks, the velocity is approximately 
linear with the slope $\inv{t}$.
\vskip 0.2cm

Shocks appear when the minimum is reached for two values $y_1$ and 
$y_2$ of $y$. Let $\Phi_{1,2}=\Phi_0(y_{1,2})$ be the value 
of the initial potential at these points. Then eq.\,(\ref{solu}) 
allows one to determine the velocity profile $u_{s}(x,t)$ around 
and inside the shocks at finite value of the viscosity $\nu$ 
by expressing $\exp\[{-\inv{2\nu} \Phi_{s}(x,t)}\]$ as the sum 
of contributions from the two minima. One obtains:
\debut
u_{s}(x,t) = \inv{t}\({x-\half(y_1+y_2)}\) - \frac{\mu_{s}}{2t} \ 
\tanh\({ \frac{\mu_{s}}{4\nu t}\({x-\xi_{s} t -\half(y_1+y_2)}\)}\) 
\label{ushock}
\fin
where $\mu_{s}=y_1-y_2>0$ and $\xi_{s}= \frac{\Phi_1-\Phi_2}{y_1-y_2}$.
The width of the shock is of order $l_c\simeq \frac{2\nu t}{\mu_s}$.
In the inviscid limit $\nu\to 0$, eq.(\ref{ushock}) becomes
\debut
u_{s}(x,t)|_{_{\nu=0}} =\, \xi_s \mp \frac{\mu_s}{2t} +\frac{x-x_s(t)}{t}
\quad {\rm for} \quad \pm(x- x_s(t))>0
\label{ushock0}
\fin
where $x_{s}(t)=\xi_{s} t +\half(y_1+y_2)$ is the time $t$ position
of the shock which moves with the velocity $\xi_{s}$ and follows 
a Lagrangian trajectory. 
The values of the velocity on the two sides of the shock are:
\debut
u_{s}^\pm \equiv u_s(x_{s}^\pm)=\xi_{s} \mp \frac{\mu_{s}}{2t} 
\label{shock+-}
\fin
so that $\mu_{s}\over t$ is the amplitude of the shock.
\vskip 0.2cm

$\bullet$ To mimic this large-time behavior, following Kida
\cite{kida}, we choose as velocity profiles the ansatz:
\debut
u(x,t)=\d_x S(x,t) \quad {\rm with}\quad
S(x,t) = \min_j\({ \phi_j + \frac{(x-y_j)^2}{2t} }\) 
\label{upoisson}
\fin

The points $(\phi_j,y_j)_{j\in {\bf Z}}$ specify a given
realization. For any realization, i.e. for
any data of the points $(\phi_j,y_j)$,  these ans\"atze (\ref{upoisson})
are solutions of the inviscid Burgers equation.
They have exact sawtooth profiles\footnote{In particular, the velocity
is not defined by the above formul\ae\ at the shocks. This is at the
origin of most of the following discussions.}
with slope $1/t$. In this ansatz all shocks are created 
at time $t=0$. The later time evolution is then governed 
by the shock collisions. Thus different times are strongly correlated.

Following Kida \cite{kida}, we shall concentrate on velocity statistics 
specified by demanding that
$(\phi_j,y_j)_{j\in {\bf Z}}$ be a Poisson point process\footnote{The
basic rules to manipulate such processes are briefly recalled in appendix
\ref{sec:app1} and \ref{sec:app2}, where some explicit computations 
are made.} with intensity $\CI = \ e^{\phi}\,d\phi\, dy$.

This choice of statistics ensures that the velocity $u(x,t)$ is
self-similar with characteristic length $l(t)\sim \sqrt{t}$ 
which means that $s\, u(sx, s^2t) \cong u(x,t)$.
Here and in the following, $\cong$ means an equality in law, 
i.e. inside any correlation functions.
We could as well choose other intensities for the
Poisson process. This amounts to choose other scalings for the
characteristic length. 

$$\epsfbox{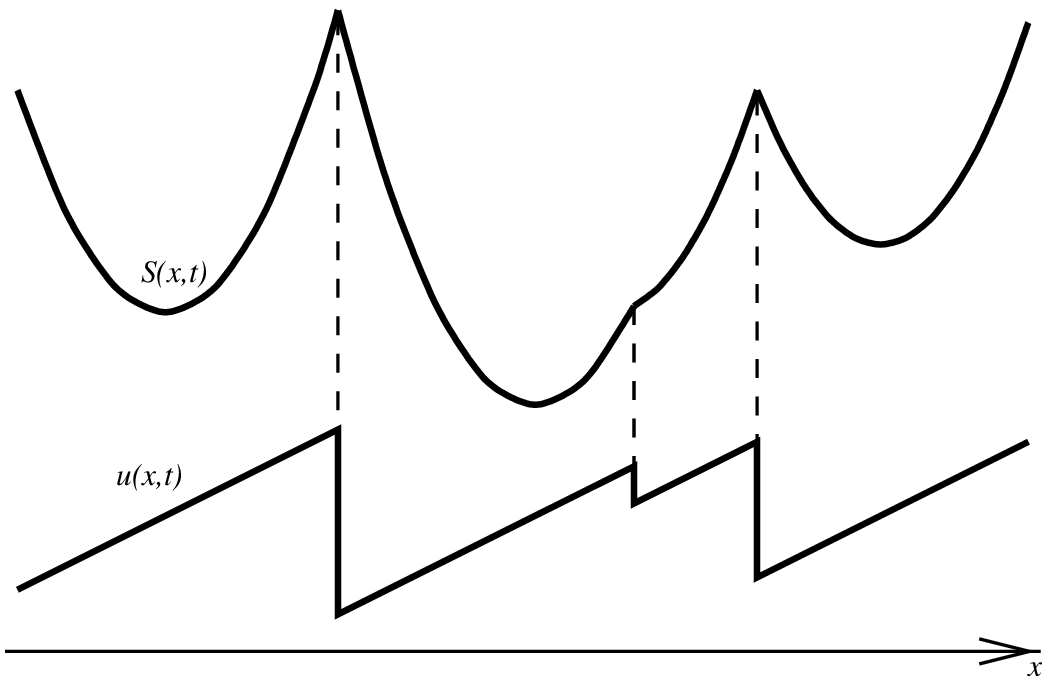}$$

\centerline{The sawtooth velocity profile.}

\section{Lagrangian trajectories and transport equations.}

Lagrangian trajectories $x(t)$ starting
at point $x_0$ at time $t_0$ are defined as solutions
of the evolution equation:
\debut
\frac{dx(t)}{dt}\ =\ u(x(t),t) 
\quad {\rm with}\quad x(t_0)=x_0 \label{trajec}
\fin
In this section, we shall specify the equation governing Lagrangian 
trajectories in the inviscid case. This requires first a detour
through the $\nu \neq 0$ situation.

$\bullet$ The above differential equation is well-posed for a velocity
field $u(x,t)$ solution of the Burgers equation
with finite non vanishing viscosity $\nu\not=0$,
since then $u(x,t)$ is smooth enough.

However the limit $\nu\to0$ is delicate:\\
--- If the point $x(t)$ of a trajectory is far from
shocks, the velocity is then regular around that point
even in the inviscid limit and the trajectory is
well defined. At large time, the velocity far from shocks
is of the form $u(x,t)=\inv{t}(x-y_*)$ with $y_*$ approximately
constant and the trajectories are then straight lines.
This applies as long as the trajectories are away from shocks.\\
--- Assuming that shocks are diluted, the trajectories near a shock in
the inviscid limit $\nu\to 0$ may be analyzed using the
velocity profile (\ref{ushock0}). In this environment, 
solutions of the Lagrange equation $\dot x= u(x,t)$ are
such that:
\debut
\sinh\({ \frac{\mu_s}{4\nu t} (x(t)-x_s(t)) }\)~ 
\exp\({-\frac{\mu_s^2}{8\nu t}}\) = {\rm constant.} \non
\fin
with $x_s(t)$ the time $t$ position of the center of the shock.
Recall that the width of the shock is of order $l_c\simeq \frac{2\nu
t}{\mu_s}$. 
This equation means that particles away from the shock take a 
finite time to enter the shock. 
Once they are in the shock they move coherently
with it with velocity almost equal to $\xi_s \equiv \dot x_s(t)$.
But they never cross the shock center.
\vskip 0.2cm

$\bullet$ If we want to recover the $\nu =0$ limit behavior directly in
the inviscid case with the ansatz (\ref{upoisson})
for the velocities, we have to be careful. At discontinuities of the
velocity, eq.(\ref{trajec}) does not make sense for two reasons :
the velocity is not defined at the shocks and
the derivative of a differentiable function cannot 
exhibit pure discontinuities.

A simple modification that will ensure the gluing of particles to
shocks, the main feature at finite but small viscosity, is the
following:\\
--- First we define $\bu(x,t) \equiv
\frac{1}{2}(u(x^+,t)+u(x^-,t))$. For the ansatz (\ref{upoisson}), this
definition makes sense for any $x$ and extends the definition of $u(x,t)$ to
shocks (obviously $u=\bu$ away from shocks).\\
--- Then we demand that trajectories be continuous and satisfy
\debut
\frac{dx(t)}{dt}\vert_+ \equiv \lim_{\ep\to 0^+} \frac{x(t+\ep)-x(t)}{\ep}= 
\bu(x,t) \label{lag0}
\fin
If we assume that the shocks form a discrete set (no limit
points)\footnote{The probability distribution for the velocities
ensures that this happens with probability one.} these two
requirements ensure that 
trajectories are uniquely defined for $t \geq t_0$ once the boundary
condition $x(t_0)=x_0$ is specified. Since the velocity of a shock is
the mean of the velocities at the points just preceding and just
following it, eq.(\ref{lag0}) ensures that particles stick to
shocks.

$\bullet$ According to the ansatz (\ref{upoisson}), away from shocks, $u(x,t)
= \inv{t}(x-y)$ for some $y$. So the trajectory
is:
\debut
x(t) = x_0 + (t-t_0) \frac{x_0- y}{t_0},
\quad {\rm away~ from~ shocks} \label{traj1}
\fin
with $x_0$ the position at time $t_0$.
This is true up to the time at which the particle meets a shock.
Shocks are at the points where
two parabol\ae\ $\phi_{1,2} + \frac{(x-y_{1,2})^2}{2t}$
minimizing eq.(\ref{upoisson}) intersect. They move
with a velocity $\xi_{12}=\frac{\phi_1-\phi_2}{y_1-y_2}$.
In the time interval during which the shock exists,
the trajectory equation is the shock equation:
\debut
x(t) = \half(y_1+y_2) + \xi_{12}\, t ,
\quad {\rm on~ the~ shock} \label{traj2}
\fin
Once a particle is on a shock it follows it and the
cascade of shocks arising from its collisions.
Note that when two shocks hit they merge into a third shock.
In particular, a particle not on a shock at time $t$ has never met a
shock before\footnote{A different proof of the same result can be
found in appendix \ref{sec:app2} where it appears as a natural part of
the argument.}. 
A general feature of the trajectories is that particles move at
constant velocity on intervals of the form $[t,t'[$ with $t < t'$. 
\vskip 0.2cm

$\bullet$ This definition of the Lagrangian trajectories ensures
the physical fact that the velocity field in the
inviscid limit is transported by the fluid. 
Indeed, since a particle moving along
a Lagrangian trajectory keeps its velocity for a finite time
interval $\ep$ with $\ep$ sufficiently small, one has:
\debut
\lim_{\ep\to 0^+}\inv{\ep}\Bigl[{
 \bu(x+\ep \bu(x,t), t+\ep) - \bu(x,t) }\Bigr] =0 \non
\fin
Accordingly, the transport equation for a tracer $T(x,t)$
moving in the inviscid velocity field $u(x,t)$ will be:
\debut
\lim_{\ep\to 0^+}\inv{\ep} \Bigl[{
 T(x+\ep \bu(x,t), t+\ep) -  T(x,t) }\Bigr]=0 \label{trans}
\fin
It coincides with the $\kappa\to 0$ limit of eq.(\ref{Tbis})
provided the Lagrangian trajectories are specified as in (\ref{lag0}).

We show now how the above equation (\ref{trans}) can be
solved. The idea is  
to find an implicit formula for the Lagrangian trajectories, taking any
number of shocks into account. Fix $x_0$ and $t_0$, and consider the
function ${\cal X}(x,t) \equiv x-x_0-(t-t_0)\bu(x,t)$ for $t \geq t_0$
and $x$ arbitrary. It is readily checked that $\frac{{\cal
X}(x,t)-{\cal X}(x',t)}{x-x'} \geq \frac{t_0}{t}$, so that for fixed
$t$, ${\cal X}(x,t)$ is a strictly increasing function of $x$ with
$\lim_{x \rightarrow \pm \infty}{\cal X}(x,t)=\pm \infty$. This means
that
we can define a function $\tilde{x}(t)$ for $t \geq t_0$ by the
condition that ${\cal X}(\tilde{x}(t)^+,t)\geq 0 \geq {\cal
X}(\tilde{x}(t)^-,t)$. It is cumbersome but straightforward to check
that $\tilde{x}(t)$ is the solution of (\ref{lag0}) with initial
condition $\tilde{x}(t_0)=x_0$. Hence the solution of (\ref{trans})
with initial condition $T(x,t_0)=\th(x-x_0)$, where $\th(x)$ is the
Heaviside step function, is
$T(x,t)=\th(x-x_0-(t-t_0)\bu(x,t))$. By linearity, the solution with
initial data $T(x,t_0)=T _0(x)$ is 
$T(x,t)=T_0(x-(t-t_0)\bu(x,t))$. This solution develops
discontinuities\footnote{But no 
nastier singularities.} at the shocks, even if the initial condition
is smooth. 

\vskip 0.2 cm

\section{Operator localized on shocks and equations of motion.}
In this section, we discuss what happens to the Burgers equation in 
the inviscid limit. We shall argue that
the actual inviscid Burgers equation is
not the naive limit $\nu\to 0$ of eq.(\ref{burgers}) but is:
\debut
\[{\, \d_tu(x,t) +\half\Bigl({u(x^+,t)+u(x^-,t)}\Bigr)\,(\d_xu(x,t))\, }\] 
\, = \,0	\label{zerobur}
\fin
with the equality valid inside correlation functions with velocity
fields (with or without derivatives)
away from $x$ and  velocity fields (without derivatives) at the point
$x$. This is not quite the usual way to 
write the inviscid Burgers equation. So we shall start with the more
familiar formul\ae\ and show the equivalence with (\ref{zerobur}). 
The argument will be based on an analysis of 
operators localized on shocks which may be used to derive
equations of motion valid inside any correlation functions.

$\bullet$ At $\nu\not= 0$ the Burgers equation (\ref{burgers})
could be written as: 
\debut
\Big(\d_t\,+\,u\d_x-\,\nu\,\d_x^2 
+\la^2\,\nu\,(\d_xu)^2\Big)\, \ee^{\lambda\, u}\,=0 \label{eqburg}
\fin
Since $\ee^{\lambda\, u}$ is finite in the inviscid regime, the
distribution $\d_x^2\ee^{\lambda\, u}$ is well defined in this limit,
and $\nu \d_x^2\ee^{\lambda\, u}$ vanishes when
$\nu\to 0$. Eq (\ref{eqburg}) can be rewritten in this limit as:
\debut
\Big(\d_tu(x,t)\,+\,u(x,t)\d_xu(x,t)\,\Big)\, \ee^{\lambda\, u(x,t)}
\,+\,\lambda\,\epsilon(x,t)\ee^{\lambda\, u(x,t)}\,\cong\, 0
\label{invbur}
\fin
Here $\ep(x,t)$ is the dissipation field defined by
$\ep(x,t) = \lim\limits_{\nu \to 0} \nu (\d_xu)^2$.
The product $u(x,t)\d_xu(x,t)$ is ill-defined since it
is a product of distributions. Eq.(\ref{invbur}) should actually
be read as:
\debut
\Big(\d_t\,+\,\la\d_{\la}\inv{\la}\d_x\,\Big)\, \ee^{\lambda\, u(x,t)}
\,+\,\lambda^2\,\epsilon(x,t)\ee^{\lambda\, u(x,t)}\,\cong\, 0
\label{invburbis}
\fin
This is the well-known inviscid equation of motion. The fact that the
dissipation field survives the inviscid limit is sometimes
called the dissipative anomaly.

\vskip 0.2cm

$\bullet$ The presence of shocks is at the origin of universal features 
which are independent of the details of the statistics. 
As explained in ref.\cite{Bbong}, they may be 
analyzed by looking at fields localized on the shocks. By definition, 
these fields may be represented for any realization as:
\debut
\CO_g (x,t) = \sum_{shocks}~ g(\xi_{s},\mu_{s})~
\de(x-x_{s}(t)) \label{loc}
\fin
where the sum is over the shocks with $x_{s}(t)$ denoting the position
of the shock, $\xi_{s}$ its velocity and $\mu_{s}\over t$ its amplitude.
The function  $g(\xi_{s},\mu_{s})$ which specifies 
$\CO_g$ will be called the form factor of the operator.

By using the velocity profile (\ref{ushock}) inside and around 
the shocks, one may map fields defined in terms of the velocity 
$u(x,t)$ into the shock representation.
The two basic examples described in ref.\cite{Bbong} 
are the generating functional 
$\({\d_x-\frac{\la}{t}}\) e^{\la\, u(x,t)}$
and $\ep(x,t)\, \ee^{\la\, u(x,t)}$
with $\ep(x,t)$ the dissipation field.
These fields are localized on the shocks. Indeed, outside
shocks $\(\d_x u(x,t)-\inv{t}\)$ vanishes 
since away from shocks, $u(x,t)=\frac{x-y_*}{t}$ with $y_*$ 
almost independent of $x$. 
Similarly, the dissipation field $\ep(x)$, which is naively zero due to
the prefactor $\nu$ in its definition, is actually a non-trivial
field since $(\d_x u)^2$ is singular in the inviscid limit.
These singularities are localized on shocks and so is the dissipation
field. In other words, dissipation takes place only at shocks. 

\vskip 0.2cm

The shock velocity profiles (\ref{ushock}) at finite $\nu$ can be used
to regularize the ill-defined expressions that arise in a na\"{\i}ve
$\nu=0$ limit. In practice, given a local functionnal of the velocity,
which is well defined at finite viscosity, one takes the $\nu =0$
limit in the distributional sense. For the above two examples, one
obtains \cite{Bbong}~:
\debut
\({\d_x-\frac{\la}{t}}\) \ee^{\la\, u(x,t)}= -2\sum_s \ee^{\la\xi_{s}}
\sinh(\frac{\la\mu_{s}}{2t})~\de(x-x_{s}(t))\,. \label{dula}
\fin
and
\debut
\ep(x,t)\, \ee^{\la\, u(x,t)}
\,=\, 2\lambda^{-3}\sum_{s} \ee^{\la\xi_{s}}\({\frac{\la\mu_s}{2t}
\cosh(\frac{\la\mu_s}{2t}) - \sinh(\frac{\la\mu_s}{2t}) }\)
\de(x-x_{s}(t))\,. \label{epla}
\fin

Now one may use the representation of the dissipation field
as an operator localized on shocks to find
alternative representations of them.
Indeed eq.(\ref{dula}) implies:
\debut
\(u({\d_xu-\inv{t}) \ee^{\la\, u} }\)_{(x,t)}
= -\frac{2}{\la^2}\sum_s \ee^{\la\xi_{s}} \({
\la\xi_s\sinh(\frac{\la\mu_{s}}{2t}) 
+\frac{\la\mu_s}{2t} \cosh(\frac{\la\mu_{s}}{2t})
- \sinh(\frac{\la\mu_{s}}{2t}) }\)~\de(x-x_{s}(t)) \non
\fin
However, looking at the product of the operator (\ref{dula})
with velocity at nearby points gives:
\debut
\half\({u(x^+,t)+u(x^-,t)}\)\({(\d_xu-\inv{t}) \ee^{\la\, u} }\)_{(x,t)}
= -\frac{2}{\la^2}\sum_s \ee^{\la\xi_{s}} 
\la\xi_s\sinh(\frac{\la\mu_{s}}{2t}) ~\de(x-x_{s}(t)) \label{avoir}
\fin
This is found using the fact that the velocity
on the two sides of the shocks are $u^\pm_s=\xi_s\mp \frac{\mu_{s}}{2t}$.
Comparing these expressions with the form factor of the dissipation
fields, eq.(\ref{epla}), gives:
\debut
\la\ep(x,t)\ee^{\la\, u(x,t)} \cong 
\({ \half\({u(x^+,t)+u(x^-,t)}\) -u(x,t)}\)
\,(\d_xu(x,t))\, \ee^{\la\, u(x,t)}
\label{eqdiss}
\fin
This is an extension of the well-known formula
$\ep(x)= \inv{12} \lim\limits_{l\to 0}\d_l\[{u(x)-u(x+l)}\]^3$.
In this formula, $u\, \d_xu\, \ee^{\la\, u}$ means
$\d_{\la}\inv{\la}\d_x\, \ee^{\lambda\, u}$. 
As expected the dissipation field is located on the discontinuity 
of the velocity field.
This relation is valid inside any correlation functions
with other fields away from point $x$.

Comparison of eq.\,(\ref{eqburg}) with eq.\,(\ref{eqdiss})
yields an alternative way of writing the inviscid
Burgers equation in which the dissipation has completely
disappeared:
\debut
\[{\, \d_tu(x,t) +\half\Bigl({u(x^+,t)+u(x^-,t)}\Bigr)\,(\d_xu(x,t))\, }\] \,
\ee^{\la\, u(x,t)} \, \cong \,0 \label{beau}
\fin
This is equivalent to eq.(\ref{zerobur}).
It has a simple interpretation: it is the simplest possible
point splitting regularization of the naive inviscid Burgers equation.
The validity of this formula can be checked by hand in simple explicit
correlation functions.
\vskip 0.2 cm

$\bullet$ Correlation functions of the velocity fields,
without any derivative, are continuous as functions of
the positions of the velocities. But the non-smoothness of the
velocities in the inviscid limit implies that correlation
functions of derivatives of the velocity field may be 
discontinuous and/or singular when points coincide.

This has echoes on the products of operators localized 
on shocks:\\
--- Products of an operator localized
on shocks times powers of the velocity field are discontinuous
at coinciding points. These properties were illustrated 
in eq.(\ref{avoir}); \\
--- Products of operators localized on shocks are singular at
coinciding points. More precisely,  
fields localized on shocks form a closed algebra \cite{Bbong}:
\debut
\CO_f(x,t)\cdot \CO_g(y,t) = \de(x-y)~ \CO_{fg}(x,t)\ 
+\ {\rm regular}\,. \label{ope}
\fin
The contact term $\de(x-y)$ in this operator product expansion arises
from the coinciding shocks in the double sum representing the product 
operator. This operator product expansion implicitly assumes that
shocks are diluted.
\vskip 0.2cm

\section{Lagrangian trajectory statistics.}

For  non time-reversal invariant velocity fields
one may consider backward and forward Lagrangian statistics.

$\bullet$ The backward statistics encodes the probability distribution
of the initial positions of the trajectories at time $t_0$
knowing their positions at later time $t>t_0$. 
For $n$-trajectories they are given by the expectation values,
\debut
P^{[n]}_{ret.}(x_j,t|x^0_j,t_0)
=\vev{\prod_{j=1}^n \CP_{ret.}(x_j,t|x^0_j,t_0)} 
\quad {\rm with}\quad
\CP_{ret.}(x,t|x^0,t_0)= \de(x^0-\hat x(t_0|x,t))
\label{pret}
\fin
with $\hat x(t_0|x,t)$ the position of the trajectory at time $t_0$
which will be at $x$ at later time $t>t_0$.

Although backward statistics are clearly a well-defined object 
from a probabilistic point of view, our representation of backward
statistics, involving $\hat 
x(t_0|x,t)$ may seem inappropriate because trajectories may
merge with increasing time, so that in general trajectories
cannot be followed for decreasing time. However, the measure
of the set for which one or more of the points $x_j$ lies exactly on a
shock at time $t$ is zero, so that the backward
trajectory is defined with probability one. This is obviously true as
long as the points $x_j$ are all distinct. When two or more of them
coincide, things are not so clear. However, we shall be able to
check explicitly that our backward statistics are well-normalized i.e. that:
\debut
\int \prod_j dx^0_j\ P^{[n]}_{ret.}(x_j,t|x^0_j,t_0) =1.\non
\fin
This ensures that we have not missed delta functions at coincident points. 

$\bullet$ The forward statistics codes the probability distribution
of the final positions of the trajectories at time $t$
knowing their initial positions at a previous time $t_0<t$.
For $n$-trajectories they 
are given by the expectation values,
\debut
P^{[n]}_{adv.}(x_j,t|x^0_j,t_0)
=\vev{\prod_{j=1}^n \CP_{adv.}(x_j,t|x^0_j,t_0) } 
\quad {\rm with} \quad 
\CP_{adv.}(x,t|x^0,t_0)=\de(x- x(t|x^0,t_0))
\label{padv}
\fin
with $x(t|x_0,t_0)$ the position of the trajectory at time $t$
which was at $x_0$ at the initial time $t_0<t$. Hence, 
$\hat x(t_0|x,t)$ and $x(t|x_0,t_0)$ are formally inverse
functions: $x(t|\hat x(t_0|x,t),t_0)=x$.
The forward probability distribution functions
 are normalized such that:
\debut
\int \prod_j dx_j\ P^{[n]}_{adv.}(x_j,t|x^0_j,t_0) =1\non
\fin

\vskip 0.2cm

$\bullet$ To deal with functions and not distributions, it is
convenient to compute expectation values of products
of step functions:
\debut
H^{[n]}(x_j,t|x^0_j,t_0)= \vev{\prod_{j=1}^n \CH(x_j,t|x^0_j,t_0)} 
\quad {\rm with}\quad
\CH(x,t|x^0,t_0) = \theta(x^0-\hat x(t_0|x,t)) \label{defH}
\fin
with $\th(z)$ the step function: $\th(z)=0$ for $z<0$ and $\th(z)=1$ for $z>0$.
The functions $H^{[n]}$ give the probabilities for particles at points
$x_j$ at time $t$ 
to be at positions above $x^0_j$ at time $t_0$.
They are such that:
\debut
P_{ret.}^{[n]}(x_j,t|x^0_j,t_0) &=& 
\prod_j\d_{x^0_j}\ H^{[n]}(x_j,t|x^0_j,t_0) \label{hret}\\
P_{adv.}^{[n]}(x_j,t|x^0_j,t_0) &=& (-)^n
\prod_j\d_{x_j}\ H^{[n]}(x_j,t|x^0_j,t_0) \label{havant}
\fin
Since $H^{[n]}(x_j,t|x^0_j,t_0)$ are expectation values of local
functional of the velocity field not involving derivatives
they can be computed directly from the velocity distribution
functions. These are recalled in Appendix A.

Remark that $P_{ret.}^{[n]}$ will be regular at coinciding points
since they do not involve derivatives of $u$, whereas $P_{adv.}^{[n]}$
will be singular since they involve such derivatives.

\vskip 0.2cm

$\bullet$ Let us now argue that the backward statistics are 
related to the joint laws of the speeds $u(x_j,t)$, at least as long
as the configuration is non-degenerate (no two points $x_j$
coincide). In this case indeed,  
with probability one, no $x_j$ lies on a shock, so each has
a speed described by a single parabola, and then the same was true at
any previous time. Hence with probability one, the particle passing at
$x_j$ at time $t$ was at $x_j-(t-t_0)u(x_j,t)$ at time $t_0$ (remember
that as long as they do not meet a shock, particles move at constant
speed). So only a trivial change of variables is needed to go from the
joint law of the initial positions $x_j^0$ to the joint law of the speeds
$u(x_j,t)$, the relation being
$u(x_j,t)=\frac{x_j-x_j^0}{t-t_0}$. This ensures that the total mass
of the backward distribution for non coincident points is unity,
so that no finite probability is carried by degenerate configurations.
This implies that 
\debut 
\CH (x,t|x^0,t_0)\ =\ \th(x^0-x+(t-t_0) u(x,t)).\non
\fin
Hence the backward probability distribution
is:
\debut
\CP_{ret.}(x,t|x^0,t_0)\ =\ \de(x^0-x+(t-t_0) u(x,t)). 
\label{pretard}
\fin
It satisfies the adequate inviscid transport equation:
\debut
\[{\, \d_t +\half\Bigl({u(x^+,t)+u(x^-,t)}\Bigr)\,\d_x \, }\]\,
\CP_{ret.}(x,t|x^0,t_0)\ \cong \ 0 \label{eqpret}
\fin
with the appropriate boundary condition:
\debut
\CP_{ret.}(x,t|x^0,t_0)\vert_{t=t_0} = \de(x-x_0). \non
\fin
Eq.(\ref{eqpret}) is valid inside correlation functions.
Note that $\CP_{ret.}$ does not satisfy the naive transport
equation (\ref{pourT}) with $\kappa=0$, since eq.(\ref{invbur})
yields:
\debut
\Bigl[{\, \d_t +u(x,t)\,\d_x\, }\Bigr]\  \CP_{ret.}(x,t|x^0,t_0)\,\cong\,
-(t-t_0)^2\, \ep(x)\, \de''(x^0-x+(t-t_0) u(x,t))\ \not= \ 0 \non
\fin
where the left hand side does not vanish due to the
dissipative anomaly.
To prove eq.(\ref{eqpret}), let us expand $\CP_{ret.}(x,t|x^0,t_0)$
in Fourier series as:
\debut
\CP_{ret.}(x,t|x^0,t_0)= \int \frac{dk}{2\pi}\ e^{ikx^0}\ 
\hat \CP_k(x,t) \quad {\rm with } \quad \hat
\CP_k(x,t)=e^{-ik(x-(t-t_0)u(x,t))}  \non
\fin
Plugging $\hat \CP_k$ into eq.(\ref{eqpret}) gives:
\debut
\[{\, \d_t +\half\Bigl({u(x^+)+u(x^-)}\Bigr)\,\d_x\, }\]\, \hat \CP_k(x,t)
&=& (ik)\[{ u(x) - \half(u(x^+)+u(x^-)) }\] \hat \CP_k(x,t) \non\\
&+& (ik(t-t_0))\[{ \d_tu(x) + 
\half(u(x^+)+u(x^-))(\d_xu(x)) }\] \hat \CP_k(x,t) \non
\fin
The first term in the r.h.s. vanishes since correlation functions of 
the velocity field without derivative are continuous whereas
the second vanishes thanks to the equation of motion (\ref{beau}).

\vskip 0.2cm

$\bullet$ The forward probability distribution is:
\debut
\CP_{adv.}(x,t|x^0,t_0) &=& 
-\d_x \CH(x,t|x^0,t_0) \non\\
&=&\Bigr({1 - (t-t_0)\d_xu(x,t)}\Bigl)\ \de(x^0-x+(t-t_0) u(x,t))
\label{pavant}
\fin
It satisfies the transport equation:
\debut
\[{\, \d_t +\d_x \,\half\Bigl({u(x^+,t)+u(x^-,t)}\Bigr)\, }\]\,
\CP_{adv.}(x,t|x^0,t_0)\ \cong \ 0 \label{eqpadv}
\fin
which corresponds to the limit $\kappa\to 0$ of eq.(\ref{rhobis}).
Remark that the Jacobian $\({1 - (t-t_0)\d_xu(x,t)}\)$ is always positive
since away from shocks $\d_xu=1/t<1/t_0$ and that on shocks 
$\d_x u$ is negative.

Eq.(\ref{pavant}) implies that the forward probability 
distribution may be decomposed as the sum
of the backward probability distribution plus an operator
which is localized on shocks. Namely:
\debut
\CP_{adv.}(x,t|x^0,t_0) = \frac{t_0}{t}\,\CP_{ret.}(x,t|x^0,t_0) -
\CD(x,t|x^0,t_0) \label{decomp}
\fin
with $\CD(x,t|x^0,t_0)=(t-t_0)\, \({ \d_xu - \inv{t}}\)\,\CP_{ret.}(x,t|x^0,t_0)$
whose shock representation is:
\debut
\CD(x,t|x^0,t_0)= 
\sum_s\ \chi\({\frac{\mu_s}{2t}\geq |\xi_s - v_{x,x_0}| }\)\
\de(x - x_s(t)) \quad {\rm with}\quad
v_{x,x_0}=\frac{x-x_0}{t-t_0}
\non
\fin
with $\chi(\CC)$ the characteristic function of the constraint $\CC$.
Here the constraint may also be written as $u^-_s\leq v_{x,x_0} \leq u^+_s$
which means that the speed of the trajectory going straight from $(x_0,t_0)$
to $(x,t)$ is between the two extreme values of the velocity at the shock.
\vskip 0.2cm

\vskip 0.2cm

\section{Lagrangian trajectory distribution functions.}

The purpose of this section is to derive explicit formul\ae\ for the
advanced and retarded one and two point function distributions of
Lagrangian trajectories. We use these results to compute the
short distance behaviour of these correlation functions,
the probability that a particle meets a shock or that two particles get
glued together. We conclude with remarks on persistence problems.  

\subsection{One-point functions.}
The one-point probability law
of the velocity field $u\equiv u(x,t)$ is:
\debut
\sqrt{\frac{t}{2\pi}} \exp\[{-\frac{tu^2}{2}}\] \ du \non
\fin
This is well known since Kida \cite{kida}, but rederived for completeness in
appendix A. Thus the one-point p.d.f. for backward and forward trajectories
coincide and are equal to:
\debut
P_{ret.}^{[1]}(x,t|x^0,t_0)=P_{adv.}^{[1]}(x,t|x^0,t_0)=
\sqrt{\frac{t}{2\pi (t-t_0)^2}}\ \exp\[{- \frac{t(x-x^0)^2}{2(t-t_0)^2}}\]
\label{onepoint}
\fin
It simply reflects the diffusion of the trajectories with
$\vev{(x-x^0)^2}\simeq (t-t_0)^2/t$. For large $t/t_0$, this is just
the ordinary dispersion of Brownian motion. But when $t-t_0$ is small
compared to $t_0$, the dispersion grows linearly with time because with
high probability no shock has been met. 

It is instructive to compare this to the probability distribution for
a particle starting at $x_0$ at time $t_0$ to flow
to $x$ at time $t$ without hitting any shock. 
As computed in Appendix B, this is equal to:
\debut
P_{{{\rm no} \atop {\rm shock}}}(x,t|x_0,t_0) \ dx=
\({\frac{t_0}{t}}\)\ \sqrt{\frac{t}{2\pi (t-t_0)^2}}\
\exp\[{ - \frac{t(x-x_0)^2}{2(t-t_0)^2} }\]\ dx \label{noshock}
\fin
In particular the probability that a particle does not meet
a shock between $t_0$ and $t$ is $t_0/t$.

The probability distribution for a particle starting at $x_0$ at time
$t_0$ to flow to $x$ at time $t$ hitting exactly $n$ shocks is more
complicated for $n \geq 0$, and it is funny that the resummations for
all values of $n$ leads to such a simple result. 
\vskip 0.2cm

\subsection{Two-point functions.}

The two-point trajectory p.d.f's are slightly more lengthy to compute.
The two-point p.d.f. for the velocity field $u_1\equiv u(x_1,t)$
and $u_2\equiv u(x_2,t)$ are recalled in appendix \ref{sec:app1}. In
the sequel, $F_t(z)$ stands for a variant of the error function defined by:
\debut
F_t(z)\ =\ e^{\frac{z^2}{2t}}\ \int^z_{-\infty}e^{-\frac{u^2}{2t}}du \non
\fin

\vskip 0.2cm

$\bullet$ Let us first look at the backward probability distribution.
Recall that it may be computed by a simple change of variables from
the velocity distribution function. 
Thus for $x_1>x_2$:
\debut
P_{ret.}^{[2]}(x,t|x^0,t_0)&=& \frac{t^2}{(t-t_0)^2} 
\de(\De-t(v_1-v_2))\frac{1}{F_t(-tv_2)+F_t(tv_1)} \non\\
&+& \frac{\De t}{(t-t_0)^2} 
\th(\De-t(v_1-v_2)) \int_{tv_1-\frac{\De}{2}}^{tv_2+\frac{\De}{2}} dz
\frac{e^{-\frac{t}{2}(v_1^2+v_2^2)}\, e^{\frac{z^2}{t}+\frac{\De^2}{4t}}}{
[F_t(\frac{\De}{2}+z) +F_t(\frac{\De}{2}-z)]^2}
\fin

\noindent
Note that as expected $P_{ret.}^{[2]}$ vanishes for 
$t(x_1^0-x_2^0) >  t_0(x_1-x_2)$ for $(x_1-x_2)>0$.
See comments below.
Note also that in the coinciding limit $x_1=x_2$ one has:
\debut
P_{ret.}^{[2]}(x,t|x^0,t_0)\vert_{x_1=x_2} &=&
\de(x_1^0-x_2^0)\ \sqrt{\frac{t}{2\pi (t-t_0)^2}}\ 
\exp\[{- \frac{t(x-x^0)^2}{2(t-t_0)^2}}\] \non\\
&=& \de(x_1^0-x_2^0)\ P_{ret.}^{[1]}(x,t|x^0,t_0) \non
\fin
This means that two trajectories at identical final positions
did start at identical initial points.
The same applies to the $n$-trajectories probability distribution
functions:
\debut
P^{[n]}_{ret.}(x_j,t|x^0_j,t_0)\,\vert_{x_n=x_{n-1}}\ =\ \de(x^0_n-x^0_{n-1})\ 
P^{[n-1]}_{ret.}(x_j,t|x^0_j,t_0) \label{biendef}
\fin
In other words, Lagrangian trajectories are statistically well-defined backwards.
\vskip 0.2cm

$\bullet$ Consider now the forward probability distribution $P^{[2]}_{adv.}$.
It is less straightforward to compute, but the relevant information can
be extracted from the formula for $H^{[2]}(x_j,t|x_j^0,t_0)$, which is
a sum of two contributions:
\debut
H^{[2]}(x_j,t|x_j^0,t_0) = K_1(x_j,t|x_j^0,t_0) + K_2(x_j,t|x_j^0,t_0)
\non
\fin
Since $H^{[2]}$ is symmetric, it is enough to evaluate it  for
$x_1>x_2$. To simplify the notations, we set 
$$\De=x_1-x_2>0 \quad {\rm  and } \quad v_j\equiv v_{x_j,x^0_j}=
\frac{x_j-x_j^0}{t-t_0}.$$
Then 
\debut
K_1(x_j,t|x_j^0,t_0)= 
\int^{\infty}_{{\rm max}(tv_1-\frac{\De}{2},tv_2+\frac{\De}{2})}
\frac{dz}{F_t(\frac{\De}{2}+z) +F_t(\frac{\De}{2}-z)} \label{k1}
\fin
and
\debut
K_2(x_j,t|x_j^0,t_0)= \frac{\De}{t}
\int^{\infty}_{tv_1} dz_1 \int^{\infty}_{tv_2} dz_2
\th (\De-(z_1-z_2))\int_{z_1-\frac{\De}{2}}^{z_2+\frac{\De}{2}} dz 
\frac{e^{-\inv{2t}(z_1^2+z_2^2)}\, e^{\frac{z^2}{t}+\frac{\De^2}{4t}}}{
[F_t(\frac{\De}{2}+z) +F_t(\frac{\De}{2}-z)]^2} \label{k2}
\fin

\vskip 0.2cm
Of course, one could recover the results for the backward
probabilities using eq.(\ref{hret}). The explicit use of
eq.(\ref{havant}) leads to formul\ae\ for the forward probability
which are not really illuminating. However, $H^{[2]}$ can be
interpreted as the probability that two particles starting at time
$t_0$ at points $x^0_1$ and $x^0_2$ respectively have absciss\ae\ at
$t$ larger than $x_1$ and $x_2$ respectively. And indeed, one can check
explicitly on the above formula for $H^{[2]}$ many expected physical
properties of trajectories:

--- Particles do not cross each other: if $\frac{x^0_1-x^0_2}{x_1-x_2} \leq
\frac{t_0}{t}$ (and in particular if $(x^0_1-x^0_2)(x_1-x_2)\leq 0$),
$H^{[2]}$ reduces to a one particle distribution:  
\debut
H^{[2]}(x_j,t|x_j^0,t_0) & = & \sqrt{\frac{t}{2\pi}} \int_{{\rm
\scriptsize max}(v_1,v_2)}^{\infty} du\, e^{-u^2t/2} \label{2en1}\\
& = & \left\{ \begin{array}{lcr} H^{[1]}(x_1,t|x_1^0,t_0) & {\rm for }
& x_1 \geq x_2 \\ H^{[1]}(x_2,t|x_2^0,t_0) & {\rm for }
& x_1 \leq x_2 \end{array}\right. .\non
\fin
Taking derivatives with respect to $x_1$ and $x_2$, one finds a
vanishing probability density if the respective orders of the
particle positions have changed between initial and final times.

--- Trajectories are well defined forward: for fixed $x_1$,$x_2$ and
$t$, formula (\ref{2en1}) is valid for $|x^0_1-x^0_2|$ small enough,
and leads to 
\debut
\lim_{x_1^0,x_2^0 \rightarrow x^0}
H^{[2]}(x_1,x_2,t|x_1^0,x_2^0,t_0)& = & H^{[1]}({\rm
max}(x_1,x_2),t|x^0,t_0)\non \\ & = & \sqrt{\frac{t}{2\pi}} \int_{\frac{{\rm
max}(x_1,x_2)-x^0}{t-t_0}}^{\infty} du\, e^{-u^2t/2}.\non
\fin
Taking the derivatives with respect to $x_1$ and $x_2$ gives:
\debut 
\label{p2p1} \lim_{x_1^0,x_2^0 \rightarrow x^0} P^{[2]}_{adv.}
(x_1,x_2,t|x_1^0,x_2^0,t_0)& = & \de(x_1-x_2)P^{[1]}_{adv.} 
(x,t|x^0,t_0) \non \\ & = & \de(x_1-x_2)\ \sqrt{\frac{t}{2\pi (t-t_0)^2}}\ 
\exp\[{- \frac{t(x-x^0)^2}{2(t-t_0)^2}}\] 
\fin 
\vskip 0.2 cm

$\bullet$ Contrary to the backward p.d.f., $P^{[2]}_{adv.}$ is singular at 
coinciding points: assuming $x^0_1 \neq x_2^0$ a direct computation
shows that
\debut
P^{[2]}_{adv.}(x_j,t|x_j^0,t_0)= R(x_1,t|x^0_j,t_0)\ \de(x_1-x_2) + \cdots 
\label{collapse}
\fin
The dots refer to terms regular at $x_1=x_2$.
The coefficient $R(x,t|x^0_j,t_0)$, which has dimension of the inverse
of a length, is the probability density of  aggregation of trajectories 
at point $x$. It is equal to:
\debut
R(x,t|x^0_1,x_2^0,t_0) =  \frac{
e^{-\frac{t}{2}(v_1^2+v_2^2)}}{2\pi t}\ 
\Bigl[{ F_t(tv_1) + F_t(-tv_2) }\Bigr] 
\quad {\rm for}\quad v_1\leq v_2 \label{tau}
\fin
Let us note that this gives also the probability that $n$ particles
have collapsed, if $v_1$ and $v_2$ refer to the speeds of the
particles with the extreme initial positions. 

This formula simplifies if one is  simply interested in the
probability that two particles starting at distinct points at $t_0$
have glued together at time $t$ : integration over the final position
gives for the total gluing probability
\debut
\frac{t-t_0}{t}\int _{\frac{|x_1^0-x_2^0|\sqrt{t}}{t-t_0}}^\infty
\frac{dv}{\sqrt{\pi}}e^{-v^2/4}.
\fin
For fixed $t_0$ and $x_1^0-x_2^0$, and $t\rightarrow \infty$ , this
behaves like $1-|x_1^0-x_2^0|/\sqrt{\pi t}-t_0/t$. This shows that
distinct particles are sure to be at the same point at a late enough
moment of the evolution, but this gluing occurs rather slowly. 

Another special case where the general formula simplifies is the limit
of identical initial positions $x^0_1=x_2^0=x^0$ leading to
\debut
R\vert_{x^0_1=x_2^0} = \sqrt{\inv{2\pi t}}\ 
\exp\[{-\frac{t(x-x^0)^2}{2(t-t_0)^2}}\] \non
\fin

This should be compared with formula (\ref{p2p1}). The difference is
exactly equal to the probability to go without shock from $x^0$ to $x$
in the time interval $[t_0,t]$ (\ref{noshock}), and this has a good
explanation: two particles starting at the same point stay sticked
together, two particles starting at distinct point may coalesce only
when they meet a shock, so the difference in collapse between starting
at the same point and starting infinitely close is simply encoded in
the probability that a single particle has met no shock. Integration
over the final points shows that the probability for two infinitely
close particle at time $t_0$ to have glued together at time $t$ is
$1-t_0/t$.
\vskip 0.2cm

The collapse probability $R$ may be computed in other ways. One way consists 
in using the operator product algebra of operator localized on shocks,
cf. eq.(\ref{ope}). Indeed, in view of the decomposition (\ref{decomp})
of $\CP_{adv.}(x,t|x^0,t_0)$, one has the following operator
product expansion:
\debut
\CP_{adv.}(x_1|x_1^0)\,\CP_{adv.}(x_2|x_2^0)
&=& \CD(x_1|x_1^0)\, \CD(x_2|x_2^0) + {\rm regular} \non\\
&=& \CR(x_1|x_j^0)\ \de(x_1-x_2) + \cdots \non
\fin
with $\CR$ the operator localized on shocks whose form factor 
is the product of those of the operator $\CD(x|x^0)$, i.e.:
\debut
\CR(x|x_j^0) = 
\sum_s \chi\({\frac{\mu_s}{2t} \geq \max_j|\xi_s-v_j|}\)\
\de(x- x_s(t)) \non
\fin
Clearly, usinf gthe shock distribution recalled in the appendix, one gets $\vev{\CR(x|x_j^0)}=R(x,t|x_j^0,t_0)$ as 
computed in eq.(\ref{tau}). 

Another way to compute $R(x,t|x_j^0,t_0)$ is as follows. We know that
particles do not cross, so that from the equation of 
trajectories, we can infer that two particles starting at $x_1^0$ and
$x_2^0$ ($x_1^0 > x_2^0$) respectively are glued together between $x$ and
$x+dx$ at time $t$ if and only if $x-x_2-(t-t_0)u(x,t) \leq 0$
(i.e. $u(x,t) \geq v_2$)  and $x+dx-x_1-(t-t_0)u(x+dx,t) \geq 0$
(i.e. $u(x+dx,t)\leq v_1 $). But as recalled in
the appendix on shock distribution functions, the probability that
$u(x,t) \geq v_2 $ and $u(x+dx,t)\leq v_1 $ for $v_2 \geq v_1$ is
\debut
\frac{dx}{2\pi}\int_{v_2}^{\infty} dv_+\int_{-\infty}^{v_1}dv_-\,
t(v_+-v_-)\th (v_+-v_-)e^{-t(v_+^2+v_-^2)/2.} 
\fin
This leads again to the above formula for $R(x,t|x_j^0,t_0)$.

\subsection{A comment on persistence problems.}

To every random velocity distribution, one can associate domains on
the $x$ axis, defined as the intervals where the velocity $v$ is
continuous. Those domains change as shocks move and annihilate into 
other shocks. This is a typical situation where persistence concepts
are useful. We have computed above two quantities that relate naturally
to persistence. For instance the probability to move on a Lagrangian 
trajectory in the time interval $[t_0,t]$ without meeting a shock,
i.e. remaining in the same domain was found to be $t_0/t$. In the same
vein, the probability for two particles starting on Lagrangian
trajectories at distance $x >0$ from each other at time $t_0$ to be at
distinct positions at time $t$ was found to be
\debut
1-\frac{t-t_0}{t}\int _{\frac{x\sqrt{t}}{t-t_0}}^\infty
\frac{dv}{\sqrt{\pi}}e^{-v^2/4},
\fin 
which behaves for large $t$ and fixed $x$ and $t_0$ as
\debut
\frac{x}{\sqrt{\pi t}}+\frac{t_0}{t}.
\fin
In particular there is no unexpected persistence exponent. 

Let us note that the more usual definition of persistence, which is
not related to Lagrangian trajectories but deals with points that
do not move with time, leads to a different kind of
behavior\footnote{This comparison was suggested to us by Claude
Godr\`eche.} that can be computed by direct use of the distribution of
velocities. For instance, the probability that a fixed point (say, the
origin) is not hit by any shock in the interval $[t_0,t]$ is 
\debut
\int dy \left[{\int dy'\,
\exp\({ \sup _{t'\in[t_0,t]}\frac{y^2-y'^2}{2t'} }\) }\right]^{-1}.
\fin
In the limit $t/t_0 \rightarrow \infty$, this exhibits the slightly
nontrivial behavior
\debut
\({\frac{2}{\pi} (\frac{t_0}{t}) \log (\frac{t}{t_0})}\)^{1/2},
\fin
quite different of the previous results for moving particles.

\vskip 0.2cm

\section{Inverse cascade.}

We now consider properties of a tracer advected in the inviscid Burgers
decaying turbulence. In
particular, we argue that there is no dissipative anomaly and that the energy
cascade is inverse. 

$\bullet$ 
As previously explained, in the inviscid limit
the appropriate transport equations are eqs.(\ref{Tbis},\ref{rhobis}).
In the limit $\kappa\to 0$, their solutions may be written in
terms of the backward and forward probability distributions.
Namely,
\debut
T(x,t) &=& \int dx^0\ \CP_{ret.}(x,t|x^0,t_0)\ T_0(x^0) \label{solT}\\
&=& T_0(x- (t-t_0)u(x,t)) \non
\fin
and
\debut
\rho(x,t) &=& \int dx^0\ \CP_{adv.}(x,t|x^0,t_0)\ \rho_0(x^0) \label{solrho}\\
&=& (1 -(t-t_0)\d_xu(x,t))\, \rho_0(x- (t-t_0)u(x,t)) \non
\fin
where $T_0(x^0)$ and $\rho_0(x^0)$ are the initial conditions
at time $t_0$.

Since correlations of the trajectory probability distributions
are computable, there are not much difficulties to evaluate
correlations of the scalars. Let us illustrate this
by showing that there is no dissipation of energy for the
tracer $T(x,t)$ and hence no dissipation anomaly for $T$.
The mechanism for that property is similar to 
the one described in ref.\cite{GaVer} in the case of the
compressible Kraichnan's model.
Assume that one is given the translation invariant two-point
function of the initial data:
\debut
\vev{T_0(x_1)T_0(x_2)} = \Ga(x_1-x_2) \non
\fin
The density of energy of the tracer is $\CE (x,t)= \half T^2(x,t)$.
Its average is:
\debut
\vev{\CE(t)} &=& 
\half \int dx_1^0dx_2^0\, P_{ret.}^{[2]}(x,x|x^0_1,x^0_2)\ 
\vev{T_0(x_1^0) T_0(x_2^0)} \non\\
&=& \half \int dx^0\, P_{ret.}^{[1]}(x|x^0)\ \Ga(0) 
=\half\ \Ga(0)\non
\fin
where we have used eq.(\ref{biendef}) for $P_{ret.}^{[2]}$
at coinciding points and the normalization condition for $P_{ret.}^{[1]}$.
Thus energy is conserved in mean, $\vev{\CE(t)}={\bar {\CE_0}}$,
and this is due to the fact that the
trajectories are statistically well-defined backward.
Notice however that at fixed initial data the density of energy
decreases at large time as $\vev{\CE (x,t)} \simeq \inv{\sqrt{2\pi t}} \int
dy T^2_0(y)$ if the integral converges.

More generally, the well-defined character of the trajectories
may also be formulated as the following operator product identity:
\debut
\CP_{ret.}(x,t|x_1^0,t_0)\, \CP_{ret.}(x,t|x_1^0,t_0)
=\ \de(x_1^0-x_2^0)\ \CP_{ret.}(x,t|x_1^0,t_0) \non
\fin
As a consequence, any products of solutions of the transport 
equation (\ref{Tbis}) at $\kappa=0$ will also be solution. 
In particular, any powers of $T(x,t)$ are also solutions:
\debut
\d_t\, T^n(x,t) + 
\half\Bigl({u(x^+,t)+u(x^-,t)}\Bigr) \d_x\, T^n(x,t) \cong 0 \non
\fin
inside correlation functions.
This shows the absence of dissipative anomalies in the
passive free advection which means that  the fields
$\kappa T^n\d_x^2 T$ vanish inside correlation
functions at $\kappa=0$.
\vskip 0.2 cm

$\bullet$ This is the sign of the absence of a direct energy cascade, 
as in the two dimensional turbulence in which the
energy cascade is inverse, i.e. toward the large scales\cite{2dtur}.
To show it more explicitly let us now assume that one is
injecting energy to the tracer such that the transport equation
is now:
\debut
\d_t T(x,t) + \half\Bigl({u(x^+,t)+u(x^-,t)}\Bigr) \d_x T(x,t)
= f(x,t) \label{forced}
\fin
with $f(x,t)$ the forcing term. 
Solutions of this equation with zero initial data at time $t_0$
are:
\debut
T(x,t) &=& \int_{t_0}^t ds \int dy\ \CP_{ret.}(x,t|y,s)\ f(y,s)
\label{solforce}\\
&=& \int_{t_0}^t ds \, f(x-(t-s)u(x,t),s) \non
\fin
Assume that the two-point function of the force is
delta-correlated in time:
\debut
\vev{f(y_1,s_1)f(y_2,s_2)} = C_L(y_1-y_2)\, \de(s_1-s_2) \label{force}
\fin
with $C_L(x)$ a smooth function varying on scale $L$ and with rapid
decrease at infinity.
The energy injection rate is $\bar e= \half C_L(0)$.
Using again the fact that trajectories are well-defined
backward, eq.(\ref{biendef}), one finds that the average
of the tracer energy density at time $t$ is:
\debut
\vev{\CE(t)} &=& \half \int_{t_0}^t ds 
\int dy\ P_{ret.}^{[1]}(x,t|y,s)\ C_L(0) \non\\
&=& \half (t-t_0)\, C_L(0)\,=\, (t-t_0)\, \bar e \label{}
\fin
where, again, we used the well-definedness of the trajectories
(see eq. (\ref{biendef})) and the normalization of the probabilities.
Thus the total amount of energy injected into the system
is transfered without dissipation. 

To decipher in which mode the energy is injected,
let us consider the scalar two-point function at
distinct points. For forcing delta-correlated in time
as in eq.(\ref{force}), the two-point function is:
\debut
\vev{T(x_2,t)T(x_1,t)}= \int_{t_0}^tdsdy_1dy_2\,
P_{ret.}^{[2]}(x_j,t|y_j,s)\ C_L(y_1-y_2) \non  
\fin
It behaves at large time and fixed positions as:
\debut
\vev{T(x_2,t)T(x_1,t)} &=& \( t\sqrt{\pi} - |x_2-x_1|\sqrt{t} \)
\int_0^1\frac{ds}{\sqrt{\pi}}\, C_L(s|x_2-x_1|) \non\\
&~& +\ F(x_2,x_1)\ +\ O(1/\sqrt{t})  \non
\fin
with $F(x_2,x_1)$ finite as $t\to \infty$
scaling as $|x_2-x_1|$ at small distance.
The energy is thus transfered to the mode corresponding to
the first line of the above equation.
Its amplitude increases with time.
It is a soft, although non constant, mode varying
smoothly and slowly.

To make manifest the absence of dissipation, consider
products of the forced scalar (\ref{solforce}) at coincident
points. One has:
\debut
T^n(x,t) = \prod_{j=1}^n \int^t_{t_0} ds_j\, f(x-(t-s_j)u(x,t),s_j) \non
\fin
Using again eq.(\ref{eqpret}) or (\ref{beau}), one deduces
that inside correlation functions:
\debut
\d_t T^n(x,t) + \half\Bigl({u(x^+,t)+u(x^-,t)}\Bigr) \d_x T^n(x,t)
\cong n\, f(x,t)\, T^{n-1}(x,t) \non
\fin
This shows that there is no dissipative anomalies at $\kappa=0$
in the scalar advection. Note that what we have described is a limit when
$\kappa$ goes to zero first and then $t$ goes to infinity to reach the
stationnary state.

Again the mechanism is similar to the one found in
compressible Kraichnan's models \cite{GaVer}: 
the injected energy is accumulated in the soft mode,
there is no dissipative anomaly
and the energy cascade is inverse. 
This is directly related to the fact that 
the trajectories are statistically well-defined. 

\vskip 0.2cm
\vskip 1.5 truecm

{\bf Acknowledgments:}
We thank K. Gawedzki and M. Vergassola for communicating us their
results \cite{GaVer} prior publication.
\vskip 0.2cm

\newpage

\appendix

\section{Velocity probability distributions.} \label{sec:app1}

In this appendix, we recall known formulae for the one point and two
point probability distributions for velocities (see eg.\cite{kida}). 
We just give a reminder of the computational rules and
illustrate it in the case of the one point velocity p.d.f. A further
illustration is given in appendix \ref{sec:app2}.

We define $S(x,t)=\min_{j}\({ \phi_j+\frac{(x-y_j)^2}{2t}}\)$ so that
$u(x,t)=\partial_x S(x,t)$. The pairs
$(\phi_j,y_j)$ are described by a Poisson point process, saying
that the cell of size $d\phi dy$ in the $(\phi,y)$-plane is occupied
with probability $e^{\phi}d\phi dy$, disjoint cells being
independent. This leads to the following useful fact that if $D$ is any
measurable set in the $(\phi,y)$-plane, the probability that all
cells in $D$ are empty is $e^{-\int _D e^{\phi}d\phi dy}
$. We call that pair $(\phi_j,y_j)$  giving the
minimum of $S$ at the point $(x,t)$ the parameters at $(x,t)$.   

\subsection{One point velocity p.d.f.}

We look for the probability $P(u(x,t) \in [v,v+dv])$. The law for the
Poisson point process reads in this case : \\
--- A cell $(\phi,y)$ with $\frac{x-y}{t} \in [v,v+dv]$ is occupied.\\
--- The cells in $D=\{(\phi',y') \mbox{ such that } \phi'+\frac{(x-y')^2}{2t} 
<\phi+\frac{(x-y)^2}{2t}=\phi+\frac{v^2t}{2}\}$ are empty.\\ 
so   
$$P(u(x,t) \in [v,v+dv])=\int_{\frac{x-y}{t} \in
[v,v+dv]}e^{\phi}d\phi dy
e^{-\int _D e^{\phi'}d\phi' dy'}.$$
Let us do this computation in detail. 
First we do the integral over $\phi'$, which varies between $-\infty$
and $\phi+\frac{(x-y)^2}{2t}-\frac{(x-y')^2}{2t}$. This yields 
$$P(u(x,t) \in [v,v+dv])=\int_{\frac{x-y}{t} \in
[v,v+dv]}e^{\phi}d\phi dy\,
e^{-\int  e^{\phi+(x-y)^2/2t-(x-y')^2/2t}dy'}.$$
Then, we integrate over $\phi$ to get
$$P(u(x,t) \in [v,v+dv])=\int_{\frac{x-y}{t} \in [v,v+dv]}
dy\,\frac{e^{-(x-y)^2/2t}}{\int
e^{-(x-y')^2/2t}dy'}.$$
Let us note that the possibility to integrate explicitly over the variable
$\phi'$ parametrizing the empty domain $D$ and over the ``center of
mass'' of the variables $\phi$ parametrizing the occupied cells is
typical. In this explicit example, the other integrations are also
immediate, but this is rather unusual.  

The $y'$ integral gives a factor $1/\sqrt{2\pi t}$ and the integration
domain for $y$ is infinitesimal, so $y=x-vt$ and $dy=tdv$.
Finally:
$$P(u(x,t) \in [v,v+dv])=\sqrt{\frac{t}{2\pi}}e^{-tv^2/2}dv.$$

Let us observe that it has total mass $1$, ensuring that this
computation, which does not take shocks into account, does not miss any
event of nonzero measure. This is a sign that shocks are diluted.

\subsection{Two point velocity p.d.f.}

We look for the probability $P(u(x_1,t) \in [v_1,v_1+dv_1],u(x_2,t)
\in [v_2,v_2+dv_2])$. By symmetry, we may (and shall) assume $x_1-x_2
\equiv \De >0$. There are two possibilities :

(*) one parabola:\\ 
--- A cell $(\phi,y)$ with $\frac{x_1-y}{t} \in [v_1,v_1+dv_1]$
and $\frac{x_2-y}{t}\in [v_2,v_2+dv_2]$ is occupied.\\
--- The cells in $D=\{(\phi',y') \mbox{ such that } \phi'+\frac{(x_1-y')^2}{2t}
< \phi+\frac{(x_1-y)^2}{2t}  \mbox{ or } \phi'+\frac{(x_2-y')^2}{2t}
< \phi+\frac{(x_2-y)^2}{2t}\}$ are empty.

(**) two parabol\ae:\\
--- A cell $(\phi_1,y_1)$ with $\frac{x_1-y_1}{t} \in [v_1,v_1+dv_1]$
and a cell $(\phi_2,y_2)$ with $\frac{x_2-y_2}{t}\in [v_2,v_2+dv_2]$
are occupied, such that $\phi_1+\frac{(x_1-y_1)^2}{2t} <
\phi_2+\frac{(x_1-y_2)^2}{2t}$ and $\phi_2+\frac{(x_2-y_2)^2}{2t} <
\phi_1+\frac{(x_2-y_1)^2}{2t}$.\\
--- The cells in $D=\{(\phi',y') \mbox{ such that } \phi'+\frac{(x_1-y')^2}{2t}
< \phi_1+\frac{(x_1-y_1)^2}{2t}  \mbox{ or } \phi'+\frac{(x_2-y')^2}{2t}
< \phi_2+\frac{(x_2-y_2)^2}{2t}\}$ are empty.

Accordingly, $P(u(x_1,t) \in [v_1,v_1+dv_1],u(x_2,t)
\in [v_2,v_2+dv_2])$ is a sum of two contributions $P_{1
\mbox{\scriptsize parab}}$ and $P_{2
\mbox{\scriptsize parab}}$ which are found after some computation to be:
\debut
P_{1 \mbox{\scriptsize parab}}= t^2dv_1dv_2 \de
(\De-t(v_1-v_2))\frac{1}{F_t(-v_2t)+F_t(v_1t)} \non
\fin 
and 
\debut
P_{2 \mbox{\scriptsize parab}}= \De t dv_1dv_2
\th (\De-t(v_1-v_2))
e^{-t(v_1^2+v_2^2)/2+\De ^2/4t} \int _{tv_1-\De/2}^{tv_2+\De/2}
dz\frac{e^{z^2/t}}{\Bigl[ F_t \left(\frac{\De}{2}-z\right)+ F_t
\left(\frac{\De}{2}+z\right) \Bigr]^2}. \non
\fin

Let us recall that $F(x)\equiv \frac{e^{x^2/2}}{\sqrt{2\pi}}\,\int_{-\infty}^x
dy e^{-y^2/2}$ and $F_t(x)\equiv \sqrt{2\pi t}F(x/\sqrt{t})$.

Again, one can check explicitly that the sum has total mass $1$, or
even better that the integral over $v_1$ or $v_2$ gives again the one
point p.d.f. This computation shows that $P_{1 \mbox{\scriptsize
parab}}$ which lives on a 
codimension one hyperplane, is completely determined as a kind of
boundary of $P_{2 \mbox{\scriptsize parab}}$. 

\subsection{Distribution of shocks.}

The two-point p.d.f for velocities allows to compute the probability
to have a shock such that  $u(x,t)=v_+$ and $u(x+dx,t)=v_-$
between $x$ and $x+dx$ by taking $\De \rightarrow 0$. The result is
\debut
\frac{dx}{2\pi}dv_+dv_-\, t(v_+-v_-)\th (v_+-v_-)e^{-t(v_+^2+v_-^2)/2.}
\fin
This can be expressed as the probability to find a shock of
amplitude $\mu/t=v_+-v_-$ and velocity $\xi=(v_++v_-)/2$ in the interval
$[x,x+dx]$ as
\debut
\frac{dx}{2\pi t}d\mu d\xi \, \mu\th (\mu)\, e^{-\xi ^2 t-\mu^2/4t.}
\fin
In particular, the probability to have a shock in the interval
$[x,x+dx]$ is $dx/\sqrt{\pi t}$. This involves only configurations with
two parabol\ae\, whereas the probability that there is
no shock in a finite interval $[x,x']$ is computed with configurations
involving one parabola and found to be $\int dy
\bigl(F_t(y-x)+F_t(x'-y)\bigr)^{-1}$. This makes it intuitively (if not
mathematically) clear that with probability one a finite interval
contains only a finite number of shocks.

\vskip 0.2cm

\section{One-point p.d.f. without shock.} \label{sec:app2}

In this appendix, we compute the probability
$P(x,t|x_0,t_0)_{\mbox{\scriptsize no shock}}$ that a particle starting
at point $x_0$ at time $t_0$ arrives in $[x,x+dx]$ at time $t$ without
ever meeting a shock.
This corresponds to the following configuration:\\ 
--- A cell $(\phi,y)$ with $x_0+\frac{x_0-y}{t_0}(t-t_0) \in [x,x+dx]$
is occupied. Let $v=(x_0-y)/t_0$.\\
--- The cells in $D=\{(\phi',y') \mbox{ such that }
\phi'+\frac{(x_0+v(t'-t_0)-y ')^2}{2t'} <
\phi+\frac{(x_0+v(t'-t_0)-y)^2}{2t'}\mbox{ for some }
t' \in [t_0,t] \}$ are empty.

The second constraint seems complicated. We claim that it is
equivalent to the extreme constraint for $t'=t$:\\
--- The cells in $D=\{(\phi',y') \mbox{ such that }
\phi'+\frac{(x_0+v(t-t_0)-y ')^2}{2t} <
\phi+\frac{(x_0+v(t-t_0)-y)^2}{2t} \}$ are empty.

This is a direct consequence of an important property of
trajectories. As already stated before, Lagrangian trajectories stick
to shocks as soon as they meet one. We can even be a bit more precise.
Suppose that at time $(x,t)$ the parabola of parameters $(\phi,y)$
dominates $(\phi',y')$, so 
\debut
\phi+\frac{(x-y)^2}{2t} < \phi'+\frac{(x-y')^2}{2t},
\fin
or better
\debut
\phi- \phi' < \frac{(y-y')(2x-y-y')}{2t}.
\fin
Consider a fictive particle moving at constant speed $v=(x-y)/t$ and
arriving at point $x$ at time $t$. At time $t_0 < t$ it was at point
$x_0=x-v(t-t_0)$. The identity
\debut
\frac{(y-y')(2x_0-y-y')}{2t_0}-\frac{(y-y')(2x-y-y')}{2t}=\frac{t-t_0}{2tt_0}
(y-y')^2 > 0
\fin  
proves that the parabola of parameters $(\phi,y)$ was already dominant
at $(x_0,t_0)$. This proves that the equivalence of the two above
definitions of the forbidden domain $D$. This means also that if point
$x$ is not on a shock at time $t$ and 
$u(x,t)=v$, there is a unique backward Lagrangian trajectory through
$(x,t)$, defined back to time $t_0$ and such that at $t_0$ the particle
was at point $x_0=x-v(t-t_0)$.

So we need to compute
$$\int_{x_0+\frac{x_0-y}{t_0}(t-t_0) \in [x,x+dx]}e^{\phi}d\phi dy
e^{-\int _D e^{\phi'}d\phi' dy'}.$$
Again, integration over $\phi'$, $\phi$ and $y'$ is straightforward,
and yields
\debut
P_{{{\rm no} \atop {\rm shock}}}(x,t|x_0,t_0) \ dx=
\({\frac{t_0}{t}}\)\ \sqrt{\frac{t}{2\pi (t-t_0)^2}}\
\exp\[{ - \frac{t(x-x_0)^2}{2(t-t_0)^2} }\]\ dx \non
\fin
Hence, the probability that no shock is met in the interval
$[t_o,t]$ is simply $t_0/t$.


\begin{thebibliography}{}
 
\bibitem{ancien} L. Richardson, Proc. Roy. Soc. London, SerA (1926) 709;\\
	A. Kolmogorov, Dokl. Akad. Nauk. 30 (1941) 301;\\
	G. Batchelor, Proc. Camb. Phil. Soc. 48 (1952) 345.

\bibitem{krach}  R. Kraichnan, Phys. Fluids 11 (1968) 945,
	 and Phys. Rev. Lett. 52 (1994) 1016.

\bibitem{incompress} See eg and refs. therein: 
	B. Shraiman, E. Siggia, C.R. Acad. Sci. 321 (1995) 279;\\
	K. Gawedzki, A. Kupiainen, Phys. Rev. Lett. 75 (1995) 3834;\\
	M Chertkov, G. Falkovich, I. Kolokolov, V. Lebedev, Phys. Rev E52 (1995) 4924 ;\\
	D. Bernard, K. Gawedzki, A. Kupiainen, J. Phys. Stat. 90 (1998) 519;\\
	O. Gat, I. Procaccia, R. Zeitak, Phys. Rev. Lett. 80 (1998) 5536;\\
	U. Frisch, A. Mazzino, A. Noullez, M. Vergassola, cond-mat/9810074.

\bibitem{compress} See eg and refs. therein:
	M Chertkov, I. Kolokolov, M. Vergassola, Phys. Rev. E 56 (1997) 5483,
	and  Phys. Rev. Lett. 80 (1998) 512.

\bibitem{GaVer} K. Gawedzki, M. Vergassola, {\it Phase transition in 
	the passive scalar advection.}, cond-mat/9811399.

\bibitem{sinaietal} W. E, K. Khanin, A. Mazel, Y. Sinai, 
	Phys. Rev. Lett. 78 (1997) 1904.

\bibitem{burg} J.M. Burgers, {\it The non-linear diffusion equation}, 
D. Reidel Publishing Co. 1974.

\bibitem{kida} S. Kida, J. Fluid. Mech. 93 (1979), 337.

\bibitem{compars} S.N. Gurbatov and A.I. Saichev, Sov. Phys. JETP 53 (1981) 347;\\
	J.D. Fournier et U. Frisch, J. M\'ec. Th\'eor. Appl. (Paris) 2 (1983) 699.\\
	 Ya. G. Sinai, Comm. Math. Phys. 148 (1992), 601.\\
	Z. She, E. Aurell and U. Frish, Comm. Math. Phys. 148 (1992), 623.\\
	M. Vergassola, B. Dubrulle, U. Frisch, A. Noullez, Astron. Atrophys. 289 (1994) 325.\\
	S. Molchanov, D. Surgailis and W. Woyczynski, Comm. Math. Phys. 168 (1995), 208.\\
	S. Gurbatov, S. Simdyankin, E. Aurell, U. Frisch and G. Toth, 
	J. Fluid. Mech. 334 (1997), 339.




\bibitem{Bbong} D. Bernard and K. Gawedzki, J. Phys. A 31 (1998) 8735.

\bibitem{2dtur} R. Kraichnan, Phys. Fluids 10 (1967) 1417 
	and J. Fluid. Mech. 47 (1971) 525.
%
\end{thebibliography}
\end{document}